\documentclass[twocolumn]{aastex63}

\usepackage{mathrsfs}

\shorttitle{Discovery of merging twin quasars at $z = 6.05$}
\shortauthors{Matsuoka et al.}

\begin{document}

\title{Discovery of merging twin quasars at $z$ = 6.05}

\correspondingauthor{Yoshiki Matsuoka}
\email{yk.matsuoka@cosmos.ehime-u.ac.jp}


\author[0000-0001-5063-0340]{Yoshiki Matsuoka}
\affil{Research Center for Space and Cosmic Evolution, Ehime University, Matsuyama, Ehime 790-8577, Japan.}

\author[0000-0001-9452-0813]{Takuma Izumi}
\affil{National Astronomical Observatory of Japan, Mitaka, Tokyo 181-8588, Japan.}

\author[0000-0003-2984-6803]{Masafusa Onoue}
\affiliation{Kavli Institute for the Physics and Mathematics of the Universe, WPI, The University of Tokyo, Kashiwa, Chiba 277-8583, Japan.}
\affiliation{Center for Data-Driven Discovery, Kavli IPMU (WPI), UTIAS, The University of Tokyo, Kashiwa, Chiba 277-8583, Japan.}
\affiliation{Kavli Institute for Astronomy and Astrophysics, Peking University, Beijing 100871, P.R.China.}

\author[0000-0002-0106-7755]{Michael A. Strauss}
\affil{Department of Astrophysical Sciences, Princeton University, Peyton Hall, Princeton, NJ 08544, USA.}

\author[0000-0002-4923-3281]{Kazushi Iwasawa}
\affil{ICREA and Institut de Ci{\`e}ncies del Cosmos, Universitat de Barcelona, IEEC-UB, Mart{\'i} i Franqu{\`e}s, 1, 08028 Barcelona, Spain.}

\author[0000-0003-3954-4219]{Nobunari Kashikawa}
\affil{Department of Astronomy, School of Science, The University of Tokyo, Tokyo 113-0033, Japan.}

\author[0000-0002-2651-1701]{Masayuki Akiyama}
\affil{Astronomical Institute, Tohoku University, Aoba, Sendai, 980-8578, Japan.}

\author[0000-0003-4569-1098]{Kentaro Aoki}
\affil{Subaru Telescope, National Astronomical Observatory of Japan, Hilo, HI 96720, USA.}

\author[0009-0007-0864-7094]{Junya Arita}
\affil{Department of Astronomy, School of Science, The University of Tokyo, Tokyo 113-0033, Japan.}

\author[0000-0001-6186-8792]{Masatoshi Imanishi}
\affil{National Astronomical Observatory of Japan, Mitaka, Tokyo 181-8588, Japan.}
\affil{Department of Astronomical Science, Graduate University for Advanced Studies (SOKENDAI), Mitaka, Tokyo 181-8588, Japan.}

\author[0000-0002-2134-2902]{Rikako Ishimoto}
\affil{Department of Astronomy, School of Science, The University of Tokyo, Tokyo 113-0033, Japan.}

\author[0000-0002-3866-9645]{Toshihiro Kawaguchi}
\affil{Department of Economics, Management and Information Science, Onomichi City University, Onomichi, Hiroshima 722-8506, Japan.}

\author[0000-0002-4052-2394]{Kotaro Kohno}
\affil{Institute of Astronomy, The University of Tokyo, Mitaka, Tokyo 181-0015, Japan.}
\affil{Research Center for the Early Universe, University of Tokyo, Tokyo 113-0033, Japan.}

\author[0000-0003-1700-5740]{Chien-Hsiu Lee}
\affil{W. M. Keck Observatory, Kamuela, HI 96743, USA}

\author[0000-0002-7402-5441]{Tohru Nagao}
\affil{Research Center for Space and Cosmic Evolution, Ehime University, Matsuyama, Ehime 790-8577, Japan.}

\author[0000-0002-0000-6977]{John D. Silverman}
\affil{Kavli Institute for the Physics and Mathematics of the Universe, WPI, The University of Tokyo, Kashiwa, Chiba 277-8583, Japan.}


\author[0000-0002-3531-7863]{Yoshiki Toba}
\affiliation{National Astronomical Observatory of Japan, Mitaka, Tokyo 181-8588, Japan.}
\affiliation{Academia Sinica Institute of Astronomy and Astrophysics, Taipei 10617, Taiwan.}
\affiliation{Research Center for Space and Cosmic Evolution, Ehime University, Matsuyama, Ehime 790-8577, Japan.}



\begin{abstract}
We report the discovery of two quasars at a redshift of $z$ = 6.05, in the process of merging.
They were serendipitously discovered from the deep multi-band imaging data collected by the Hyper Suprime-Cam (HSC) Subaru Strategic Program survey.
The quasars, HSC $J$121503.42$-$014858.7 (C1) and HSC $J$121503.55$-$014859.3 (C2), both have luminous ($>$10$^{43}$ erg s$^{-1}$) Ly$\alpha$ emission with a clear broad component (full width at half maximum $>$1000 km s$^{-1}$).
The rest-frame ultraviolet (UV) absolute magnitudes are $M_{1450} = -23.106 \pm 0.017$ (C1) and $-22.662 \pm 0.024$ (C2).
Our crude estimates of the black hole masses provide $\log (M_{\rm BH}/M_\odot) = 8.1 \pm 0.3$ in both sources.
The two quasars are separated by 12 kpc in projected proper distance, bridged by a structure in the rest-UV light suggesting that they are undergoing a merger.
This pair is one of the most distant merging quasars reported to date, providing crucial insight into galaxy and black hole build-up in the hierarchical structure formation scenario.
A companion paper will present the gas and dust properties captured by Atacama Large Millimeter/submillimeter Array observations, 
which provide additional evidence for and detailed measurements of the merger and also demonstrate that the two sources are not gravitationally-lensed images of a single quasar.


\end{abstract}




\section{Introduction} \label{sec:intro}

Quasars at high redshifts are an important and unique probe of the epoch of reionization (EoR, referring to $z \ge 6$ in this letter), 
a critical epoch for understanding the seeding and initial growth of supermassive black holes (SMBHs), the evolution of the host galaxies at an early stage of hierarchical structure formation, 
and the spatial and temporal progress of the reionization.
A significant number of EoR quasars have been discovered in the past few decades, exploiting wide-field ($>$100-deg$^2$ class) optical and near-infrared (IR) imaging surveys 
\citep[e.g.,][and references therein]{fan23}.
The ultraviolet (UV) quasar luminosity function (QLF) has now been established at $z = 6$ and $7$ \citep[e.g.,][]{p5,p19,schindler23}, demonstrating that UV emission from quasars makes only a minor contribution to cosmic reionization, if the flat QLF slope below the characteristic luminosity continues to the unobserved faint end.

In the meantime, the {\it James Webb Space Telescope (JWST)} has produced groundbreaking results in the past two years. 
Its extremely high IR sensitivity has revealed a weak broad component in Balmer emission lines of many high-$z$ galaxies, signaling the presence of low-luminosity active galactic nuclei (AGNs) 
out to $z \sim 11$ \citep[e.g.,][]{greene23, maiolino23}.
The number density of such AGNs exceeds the extrapolation of the classical QLF by several orders of magnitude \citep[e.g.,][]{harikane23, matthee23}, 
changing the paradigm of SMBH activity happening in the EoR.

On the other hand, there are still missing pieces in the AGN demographics in the EoR, one of which is pairs of quasars or AGNs in mergers.
Hierarchical structure formation within the Lambda Cold Dark Matter model suggests that galaxies grow via frequent mergers.
If a significant fraction of those galaxies contain a SMBH at the center, as implied from the measurements in the local Universe \citep[e.g.,][]{kormendy13}, then one would naturally 
expect SMBH pairs in the merging galaxies.
If the merger induces gas inflow toward the SMBHs \citep[e.g.,][]{hopkins06}, then such systems would be observed as pairs of quasars or AGNs.
The observed frequency of such pairs constrains many key factors, such as the relative importance of mergers for galaxy and SMBH evolution, the timescales associated with SMBH interaction and 
coalescence, and the number density of possible gravitational wave sources.
Quasar pairs can also serve as a signpost of galaxy overdense regions \citep[e.g.,][]{onoue18} and as a probe of the small-scale distribution of the foreground intergalactic medium \citep{rorai17}.

Searches for pairs of quasars or AGNs have used various techniques, typically based on wide-field surveys \citep[e.g.,][and references therein]{derosa19}.
The most recent efforts include those reported by \citet{silverman20} and \citet{tang21}, who used Subaru Hyper Suprime-Cam (HSC; see below) high-resolution images and identified pairs at $z \le 3$
among the known Sloan Digital Sky Survey \citep[SDSS;][]{york00} quasars.
\citet{shen21} exploited astrometry information from the {\it Gaia} satellite mission \citep{gaia16} to find two pairs among the SDSS quasars at $z = 2 - 3$.
On the other hand, no pairs were found by \citet{sandoval23} in their search from a large X-ray catalog at $z \sim 3$.
There are also projects to search for quasar pairs motivated by investigation of gravitational lensing \citep[e.g.,][]{richards06, inada12, yue23}.
From such a project based on the Dark Energy Survey \citep{abbott18}, 
\citet{yue21} found a candidate of quasar pair at $z = 5.66$, with a separation of 7.3 kpc. 
If confirmed, this would be the first quasar pair reported in the EoR.
In addition, quasars with merging galaxy companions have been reported in the EoR \citep[e.g.,][]{decarli17, decarli19b}, in which the companion galaxies are frequently invisible in the rest-UV and are only identified by submillimeter observations.
Most recently, {\it JWST} observations are finding signatures of dual AGNs in individual EoR galaxies, via double components or off-nucleus emission of broad Balmer lines \citep{maiolino23, ubler23}.

This letter presents the discovery of a pair of merging quasars at $z$ = 6.05, HSC $J$121503.42$-$014858.7 and HSC $J$121503.55$-$014859.3 (C1 and C2, hereafter).
The two quasars are separated by 12 kpc, forming one of the most distant pairs of quasars or AGNs reported to date.
We describe the target selection and spectroscopic observations in \S \ref{sec:obs}.
The nature of the two sources  is discussed in \S \ref{sec:results}, based on their imaging and spectroscopic properties.
A summary appears in \S \ref{sec:summary}.
We adopt the cosmological parameters $H_0$ = 70 km s$^{-1}$ Mpc$^{-1}$, $\Omega_{\rm M}$ = 0.3, and $\Omega_{\rm \Lambda}$ = 0.7.
All magnitudes refer to CModel magnitudes from the HSC data reduction pipeline, which are measured by fitting galaxy models convolved with the point spread function (PSF) to the observed source profile \citep{bosch18}.
The magnitudes have been corrected for Galactic extinction \citep{schlegel98}, and are reported in the AB system \citep{oke83}. 
A companion paper (T. Izumi et al., in prep.) will present the gas and dust properties of these quasars captured by Atacama Large Millimeter/submillimeter Array (ALMA) observations, as well as their kinematic modeling.

\section{Observations} \label{sec:obs}

\begin{figure*}
\epsscale{1.1}
\plotone{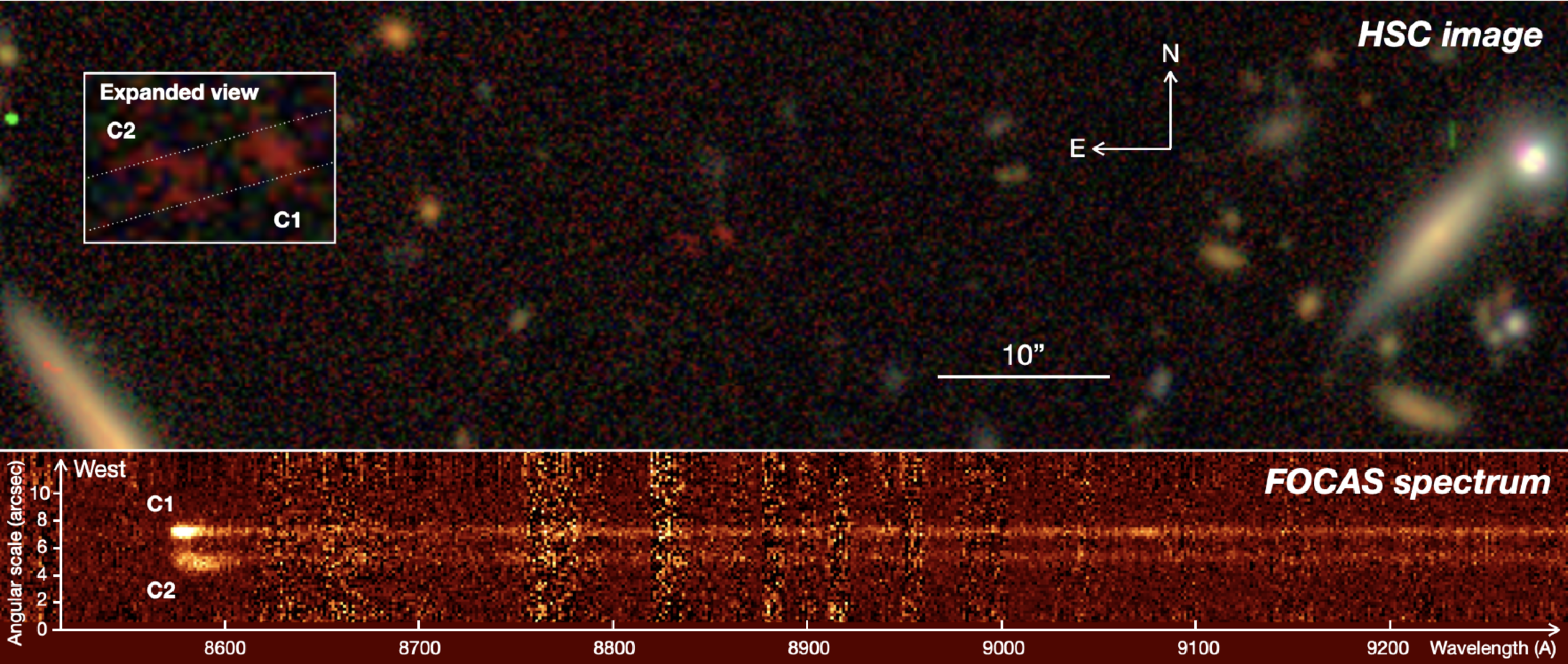}
\caption{Top: three-color (HSC $r$, $i$, and $z$-band) composite image around C1 and C2, the two reddest sources at the center.
North is up and East to the left, and the image size is approximately 90\arcsec $\times$ 25\arcsec. 
The limiting magnitude for point sources is $\sim$26.
The inset shows an expanded view of C1 and C2, with the thin dotted lines representing a 1\arcsec.0 slitlet used for FOCAS spectroscopy.
Bottom: two-dimensional FOCAS spectrum of C1 (upper trace of light) and C2 (lower trace), created by stacking all available data.
\label{fig:fc}}
\end{figure*}

Figure \ref{fig:fc} presents a three-color (HSC $r$, $i$, and $z$-band) composite image around the two quasars, C1 (West) and C2 (East).
Their observed properties 
are summarized in Table \ref{tab:measurements}.
Here $\mu_{z/y}$ represents the second-order moment of the source on the $z/y$-band image, normalized to those of field stars as a model of PSF (i.e., an unresolved source has $\mu_{z/y} = 1$).
C1 was originally selected from the HSC Subaru Strategic Program \citep[SSP;][]{aihara18} imaging survey.
Its red $i - z$ and relatively blue $z - y$ colors as well as the fact that it is not (or only marginally) spatially resolved made it an EoR quasar candidate 
in our ``Subaru High-$z$ Exploration of Low-Luminosity Quasars (SHELLQs)'' project \citep{p1,p4,p2,p5,p10,p7,p16,p19}.\footnote{
We clarify that the present two quasars were not included in the previous SHELLQs publications, and are reported here for the first time.}
The initial follow-up spectroscopy was carried out with Subaru Telescope on April 24, 2018, as a part of the Subaru intensive program S16B-011I. 
We used the Faint Object Camera and Spectrograph \citep[FOCAS;][]{kashikawa02} in the multi-object spectroscopy mode.
The combination of the VPH900 grism, SO58 order-sorting filter, and 1\arcsec.0 slitlets yielded spectral coverage from 0.75 to 1.05 $\mu$m with resolution $R \sim 1200$.
The slit angle\footnote{Slit angle is measured from North to East, such that 90$^\circ$ refers to a slit aligned to the East -- West direction.
} was set to 90$^\circ$.
We took seven 10-min exposures under the clear sky, with the seeing conditions of 0\arcsec.8 -- 1\arcsec.0.
The data reduction was performed with the Image Reduction and Analysis Facility (IRAF) using the dedicated FOCASRED package in a standard manner.
The wavelength scale was calibrated with reference to sky emission lines, and the flux calibration was tied to Feige 34, a white dwarf standard star, observed on the same night.
Slit losses were corrected for by scaling the spectrum to match the HSC $z$-band magnitude.

\begin{deluxetable*}{cccccccc}
\tablecaption{Imaging and spectroscopic measurements \label{tab:measurements}}
\tablehead{
} 
\startdata
\colhead{Object} & \colhead{R. A.} & \colhead{Decl.} & \colhead{$g_{\rm AB}$} & \colhead{$r_{\rm AB}$} & \colhead{$i_{\rm AB}$} & \colhead{$z_{\rm AB}$} & \colhead{$y_{\rm AB}$}\\\hline
C1 & 12:15:03.42 & $-$01:48:58.7 & $26.25 \pm 0.28$ & $<26.09$ & $25.73 \pm 0.22$ & $23.78 \pm 0.11$ & $23.14 \pm 0.12$ \\
C2 & 12:15:03.55 & $-$01:48:59.3 & $<26.75$              & $<26.32$ & $<26.50$              & $24.40 \pm 0.15$ & $23.75 \pm 0.18$ \\\hline
\colhead{} &  \colhead{$\mu_z$ (HSC)} & \colhead{$\mu_y$ (HSC)} &  \colhead{$\mu_z$ (FOCAS)} & \colhead{$\mu_y$ (FOCAS)}\\ \hline
C1 & 1.35 $\pm$ 0.16 & 1.27 $\pm$ 0.15 & $1.29 \pm 0.16$ & $1.34 \pm 0.15$\\
C2 & 1.60 $\pm$ 0.20 & 0.99 $\pm$ 0.23 & $1.92 \pm 0.25$ & $1.63 \pm 0.21$ \\\hline
\colhead{} & \colhead{$z_{\rm Ly\alpha}$} & \colhead{$M_{1450}$}  & \colhead{$L_{\rm bol}$ (erg s$^{-1}$)} & \colhead{$v_{\rm FWHM}$ (km s$^{-1}$) } & \colhead{EW$_{\rm rest}$ (\AA)}  & \colhead{$L_{\rm line}$ (erg s$^{-1}$)} & \colhead{Comment}\\\hline
            C1 & 6.053 &  $-23.106 \pm 0.017$           &   $(6.2 \pm 0.1) \times 10^{45}$ &   &    &   &  \\
                  &          &                                                &   &1450 $\pm$ 170   & 15 $\pm$ 2   & $(1.34 \pm 0.13) \times 10^{43}$   & Ly$\alpha$ (broad)\\
                  &          &                                                &   & 360 $\pm$ 30      &  12 $\pm$ 2   & $(1.02 \pm 0.12) \times 10^{43}$  & Ly$\alpha$ (narrow)\\
            C2 & 6.053 &   $-22.662 \pm 0.024$           &    $(4.1 \pm 0.1) \times 10^{45}$ &   &    &    \\
                  &          &                                                &   & 1290 $\pm$ 60  & 27 $\pm$ 2   & $(1.84 \pm  0.07) \times 10^{43}$  & Ly$\alpha$ (broad)\\
\enddata
\tablecomments{The magnitude lower limits are given at 3$\sigma$ confidence level.
The line FWHMs ($v_{\rm FWHM}$) have been corrected for line broadening due to the finite instrumental resolution.
The equivalent widths (EW$_{\rm rest}$) are reported in the rest frame. }
\end{deluxetable*}

The initial spectroscopy revealed strong and asymmetric Ly$\alpha$ emission at the observed wavelength of $\lambda_{\rm obs} = 8576$ \AA,  
indicating that C1 exists at $z_{\rm Ly\alpha} = 6.053$.
Soon after the spectroscopic identification, we noticed that C1 is accompanied by a fuzzy source with similar $i - z$ and $z - y$ colors (see Figure \ref{fig:fc}).
This fuzz, named C2, is separated by 2\arcsec.0 from C1 toward the East.
We carried out another set of spectroscopy with FOCAS on April 25, 26, and May 10, 2019, as a part of the Subaru intensive program S18B-071I.
This time we oriented the slit angle to 106$^\circ$, so that C1 and C2 were observed simultaneously. 
The total exposure time in this run was 270 min.
The sky condition was mostly clear, with the seeing of 0\arcsec.4 -- 0\arcsec.7. 
All the other instrument configurations and data reduction were identical to those in the initial spectroscopy.
We further obtained additional exposures totaling 100 min on March 2, 2021, 
using the same observational settings as in the 2019 run.
The sky condition was clear with the seeing of 0\arcsec.6.

We also acquired near-IR spectra of the two sources with the Fast Turnaround program (ID: GN-2020A-FT-106) at the Gemini North telescope.
We used the Gemini Near-InfraRed Spectrograph \citep[GNIRS;][]{elias06} in the cross-dispersed mode, with the 32 l/mm grating and the central wavelength set to 1.65 $\mu$m.
The slit width was 1\arcsec.0, giving spectral coverage from 0.85 to 2.5 $\mu$m and resolution $R \sim 500$.
We oriented the slit angle to 106$^\circ$ and took 63 $\times$ 5-min exposures in total, spread over a month (June 3, 4, 14, July 6, and 7, 2020). 
The observations were carried out in the queue mode, with the requested sky conditions of 50-percentile cloud coverage and 70-percentile image quality.
The data reduction was performed with IRAF using the Gemini GNIRS package in a standard manner.
The wavelength scale was calibrated with reference to Argon lamp spectra.
The flux calibration and telluric absorption correction were tied to standard stars HIP 54849 and HIP 61637, 
observed immediately before or after the target observations at similar airmass.
We scaled the GNIRS spectrum to match the FOCAS spectrum where they overlap in wavelength.

\section{Results and discussion} \label{sec:results}

\subsection{Nature of the two sources} 

\begin{figure}
\epsscale{2.3}
\plottwo{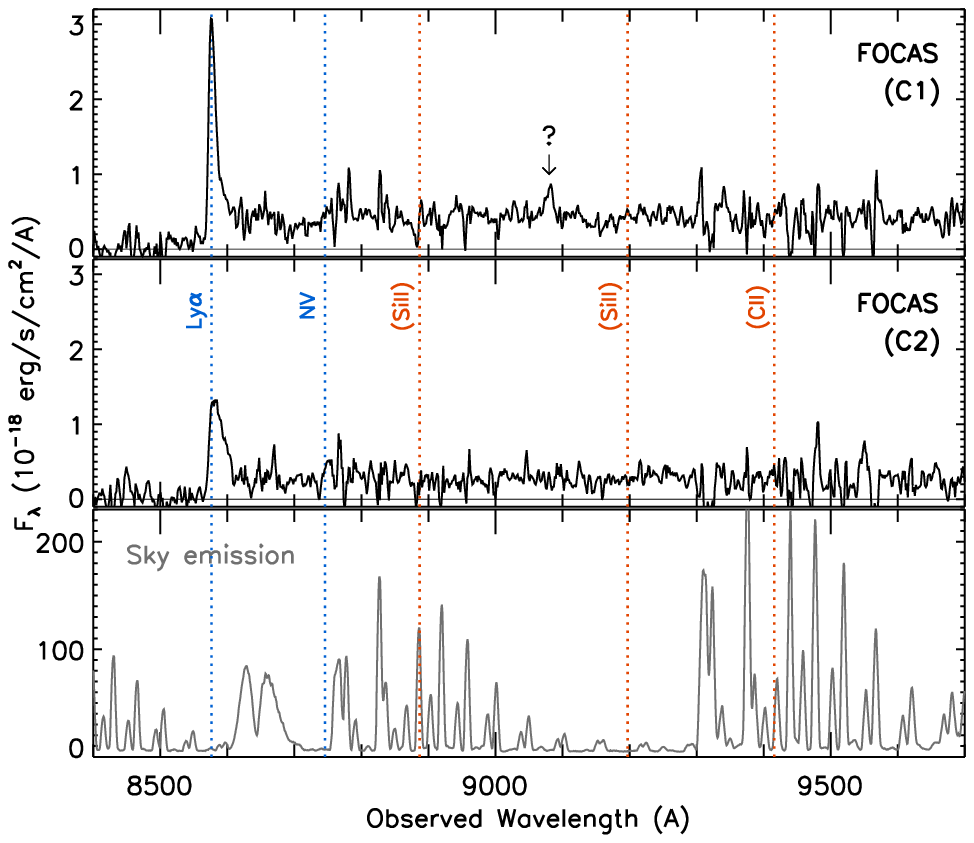}{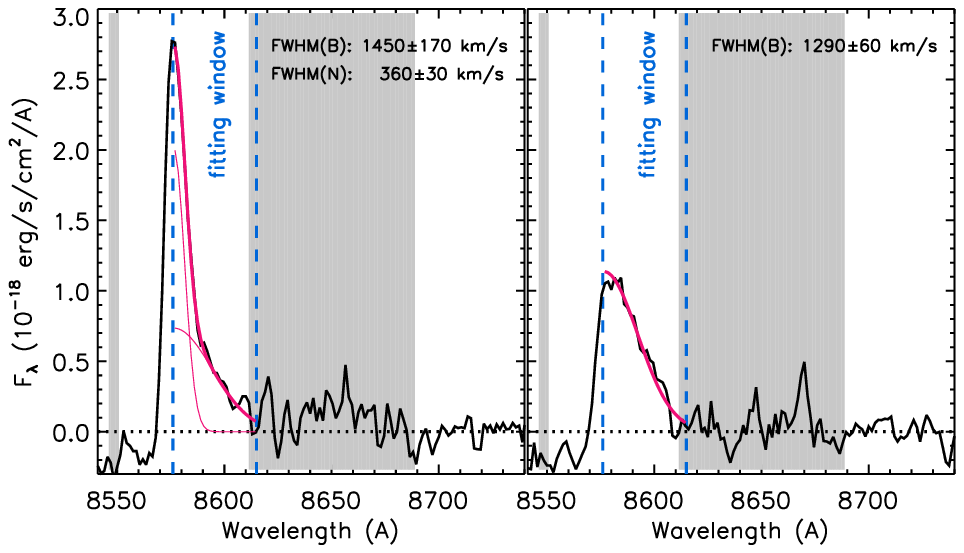}
\caption{
{\it Upper panels:} FOCAS spectra of C1 (top) and C2 (middle) created by stacking all available data, along with a sky spectrum as a guide to the expected noise (bottom). 
The dotted lines represent the expected positions of Ly$\alpha$ and \ion{N}{5} $\lambda$1240 emission lines, as well as interstellar absorption lines of
\ion{Si}{2} $\lambda$1260, \ion{Si}{2} $\lambda$1304, and \ion{C}{2} $\lambda$1335, given the redshift of $z_{\rm Ly\alpha} = 6.053$.
An unidentified line at $\lambda_{\rm obs}$ = 9082 \AA\ in C1 (see text) is marked by an arrow.
{\it Lower panels:} Continuum subtracted spectra of C1 (left) and C2 (right) around Ly$\alpha$. 
The thick red lines represent the best-fit models, while the thin red lines (only in C1) represent their broad and narrow components.
The spectral window used for the fitting ($\lambda_{\rm obs} = 8576 - 8615$ \AA) is shown by the vertical dashed lines.
The gray shaded area marks the wavelength range affected by strong sky emission.
The spectra in both upper and lower panels were smoothed using inverse-variance-weighted means over 3 pixels, for display purposes.
\label{fig:spec1}}
\end{figure}

Figure \ref{fig:fc} presents the two-dimensional FOCAS spectra of C1 (West) and C2 (East), co-added across all exposures.
The extracted one-dimensional spectra are shown in Figure \ref{fig:spec1} (upper panels).
We detected a strong emission line also from C2, whose peak wavelength is consistent with that measured in C1. 
The asymmetric profiles and the presence of the \citet{gunn65} trough at the shorter wavelengths confirm the identification of the line as Ly$\alpha$, redshifted to $z_{\rm Ly\alpha} = 6.053$.
Since Ly$\alpha$ redshifts of EoR objects are relatively uncertain (see also below), we report a formal redshift of $z = 6.05$ in this paper.
The full widths at half maximum (FWHM) of Ly$\alpha$, uncorrected for intergalactic medium (IGM) absorption, are $v_{\rm FWHM}$ = 320 $\pm$ 20 km s$^{-1}$ and 
810 $\pm$ 180 km s$^{-1}$ for C1 and C2, respectively.
We also detected flat continuum emission redwards of the line. 
The rest-UV absolute magnitudes of the two sources are 
$M_{1450} = -23.106 \pm 0.017$ (C1) and $-22.662 \pm 0.024$ (C2) at the rest-frame wavelength $\lambda_{\rm rest}$ = 1450 \AA.
These values were obtained by extrapolating the continuum flux density at $\lambda_{\rm obs} = 9000 - 9300$ \AA,
where the sky emission is relatively weak, with a power-law model with a slope $\alpha = -1.5$ \citep[$F_\lambda \propto \lambda^{-1.5}$; e.g.,][]{vandenberk01}.
Assuming a quasar bolometric correction of BC$_{1350}$ = 3.81 \citep{shen11}, we get the bolometric luminosity of $L_{\rm bol}$ = (6.2 $\pm$ 0.1) $\times$ 10$^{45}$ erg s$^{-1}$ and (4.1 $\pm$ 0.1) $\times$ 10$^{45}$ erg s$^{-1}$ for C1 and C2, respectively.

It is clear from Figure \ref{fig:spec1} (upper panels) that the Ly$\alpha$ profile
has a relatively broad component in both sources, with a narrow core component seen only in C1.
We fit two Gaussians and one Gaussian to the C1 and C2 spectra redward of the line peak, respectively,
as displayed in Figure \ref{fig:spec1} (lower panels).
The local continuum emission was estimated at $\lambda_{\rm obs} = 8694 - 8738$ \AA, where strong lines from the targets and the sky are absent,
and was subtracted before the model fitting.
We found that the broad Ly$\alpha$ components of the two sources have similar widths, 
$v_{\rm FWHM}$ = 1450 $\pm$ 170 km s$^{-1}$ (C1) and 1290 $\pm$ 60 km s$^{-1}$ (C2).
Luminosity and other line properties from the best fit models are reported in Table \ref{tab:measurements}.

While the redshift was fixed to $z_{\rm Ly\alpha} = 6.053$ during the model fitting, adopting alternative values doesn't change our conclusion that a broad line component is present.
Due to severe absorption from the IGM, the intrinsic peak of Ly$\alpha$ is often located blueward of the observed peak, which would indicate an intrinsically broader line width
than estimated above.
This is likely the case for C2, whose ALMA observations of the [\ion{C}{2}] 158 $\mu$m line indicate $z_{\rm [C II]} = 6.044$ (T. Izumi et al., in prep.).
On the other hand, C1 has $z_{\rm [C II]} = 6.057$, i.e.,
the observed Ly$\alpha$ peak is blueshifted relative to [\ion{C}{2}].
When we fix the Ly$\alpha$ redshift of the broad component to $z_{\rm [C II]} = 6.057$, we get its width of $v_{\rm FWHM}$ = 1100 $\pm$ 100 km s$^{-1}$ in C1 and 1100 $\pm$ 60 km s$^{-1}$ in C2.
We also note that the FWHM estimate remains almost unchanged when we fit only the nuclear part of the spatially resolved C2 spectrum.

The spectral properties mentioned above suggest the presence of quasars in both sources.
The widths of the Ly$\alpha$ broad components exceed the common threshold of quasar classification, $v_{\rm FWHM}$ = 500 -- 1000 km s$^{-1}$ \citep[e.g.,][]{schneider10, paris12},
and are similar to values found in faint AGNs revealed by {\it JWST} spectroscopy of EoR galaxies \citep[e.g.,][]{greene23, harikane23, maiolino23}.
These values are found at the lower end of the FWHM distribution of low-$z$ Seyfert 1 galaxies in SDSS \citep[e.g.,][]{hao05} and of high-$z$ low-luminosity quasars found in SHELLQs.
On the other hand, star-forming galaxies cannot produce line components that are significantly broader than $\sim$500 km s$^{-1}$, even with outflows \citep[e.g.,][]{newman12, swinbank19}.
We note that Ly$\alpha$ is spatially resolved in C2, and the line component extending to $>$1000 km s$^{-1}$ belongs to the nuclear part of the two-dimensional spectrum (see also below).
The observed Ly$\alpha$ luminosities of $>$10$^{43}$ erg s$^{-1}$ are also very high for non-AGN galaxies, and overlap with the lower end of the distribution of other SHELLQs quasars \citep[e.g.,][]{onoue21}.
At lower redshifts ($z \sim 2 - 3$), Ly$\alpha$ emitters with such high luminosities almost always harbor AGNs, identified via
characteristic X-ray, UV, radio continuum emission and/or high-ionization optical lines \citep[e.g.,][]{konno16, sobral18, spinoso20}.
The continuum luminosities of C1 and C2 ($M_{1450} \sim -23$ mag) are roughly 10 times higher than the characteristic luminosity of the galaxy luminosity function at $z = 6$ \citep{harikane22},
and it would be unexpected (though not impossible) if a close pair of such luminous high-$z$ galaxies were found.

\begin{figure}
\epsscale{1.2}
\plotone{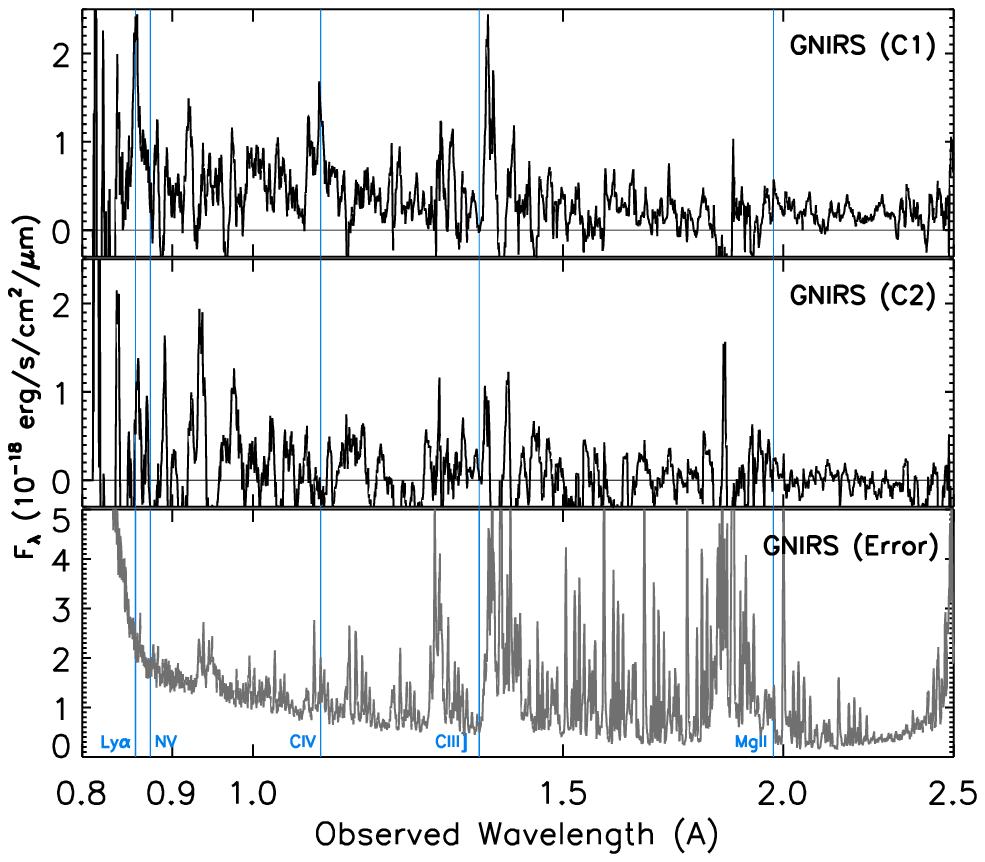}
\caption{GNIRS spectra of C1 (top) and C2 (middle) created by stacking all available data, along with the error spectrum dominated by the sky background (bottom). 
The vertical lines represent the expected positions of Ly$\alpha$, \ion{N}{5} $\lambda$1240, \ion{C}{4} $\lambda$1549, \ion{C}{3}] $\lambda$1906, and \ion{Mg}{2} $\lambda$2800, given the redshift of $z_{\rm Ly\alpha} = 6.053$.
The spectra were smoothed using inverse-variance-weighted means over 9 pixels, for display purposes.
\label{fig:spec3}}
\end{figure}

Other than Ly$\alpha$, no strong emission lines are detected from C1 or C2.
We found a small spectral bump at the expected wavelength of \ion{N}{5} $\lambda$1240 in both C1 and C2 (see Figure \ref{fig:spec1}), but the adjacent bright sky emission
hampers robust identification of this feature. 
The 3$\sigma$ upper limit of the \ion{N}{5}/Ly$\alpha$ (broad) ratio is $\sim$0.2 in both sources, which 
is consistent with the ratio measured in low-$z$ SDSS quasars \citep[$\sim$0.02;][]{vandenberk01}.
The GNIRS spectra of the two targets are very noisy (see Figure \ref{fig:spec3}) even with $>$5-hour on-source exposure, only allowing us to identify continuum emission from C1.
There is a spectral bump at the expected position of \ion{C}{4} $\lambda$1549 in C1, but the detection is marginal at most.
On the other hand, the optical spectrum of C1 (Figure \ref{fig:spec1}) exhibits a weak but clear emission line at $\lambda_{\rm obs} = 9082$ \AA, which is also apparent in the two-dimensional spectrum 
in Figure \ref{fig:fc}.
This line corresponds to $\lambda_{\rm rest} = 1288$ \AA\ at $z_{\rm Ly\alpha} = 6.053$, where no emission line is known.
It could be due to an overlapping foreground source, whose faint blue emission extends Northward of C1 (see the HSC image of Figure \ref{fig:fc}), but the present data cannot provide
any robust identification.

It is well known that quasar emission line properties, in particular those of \ion{C}{4} $\lambda$1549, \ion{Mg}{2} $\lambda$2800, and H$\beta$, are sensitive to SMBH masses ($M_{\rm BH}$).
Correlation in the form of $M_{\rm BH} \propto v_{\rm FWHM}^2 L_{\rm line}^\gamma \equiv \mathscr{M}_{\rm line}$ is observed for the above three lines,
where $L_{\rm line}$ is the line luminosity and $\gamma$ is a constant close to 0.5 \citep[e.g.,][]{vestergaard06}.
Here we obtain crude mass estimates of the two quasars via the broad Ly$\alpha$ component, which is also sensitive to $M_{\rm BH}$ \citep[e.g.,][]{takahashi24}.
As is clear from Table \ref{tab:measurements}, C1 and C2 have similar Ly$\alpha$ properties in the broad components, suggesting similar $M_{\rm BH}$. 
We looked into the spectroscopic properties of SDSS quasars measured by \citet{rakshit20}, and found 678 / 579 quasars whose $\mathscr{M}_{\rm Ly\alpha}$ ($\gamma = 0.5$ is assumed) values lie within $\pm$0.1 dex of the Ly$\alpha$ broad component of C1 / C2. 
Both of these matched samples have median masses $\log (M_{\rm BH}/M_\odot) = 8.1$, with a relatively small scatter of 0.3 dex.
We thus estimate that both C1 and C2 have $\log (M_{\rm BH}/M_\odot) = 8.1 \pm 0.3$.
The corresponding Eddington ratios are $\sim$0.4 and $\sim$0.3 for C1 and C2, respectively.
These estimates are approximate at most, and need to be updated with future measurements of, e.g., Balmer lines in the rest-frame optical with {\it JWST}.


Similar objects with luminous ($>$10$^{43}$ erg s$^{-1}$) and relatively narrow (total FWHM of $< 500$ km s$^{-1}$, uncorrected for IGM absorption) Ly$\alpha$ have been identified at $z \ge 6$ in our SHELLQs project \citep[e.g.,][]{p16}.
Though most of them do not exhibit a broad Ly$\alpha$ component, we consider them be candidate (possibly obscured) quasars, based on their high Ly$\alpha$ luminosities.
We have carried out follow-up IR spectroscopy of seven such objects with $L_{\rm Ly\alpha} = 10^{43.3} - 10^{44.3}$ erg s$^{-1}$, and found positive evidence for the presence of AGN overall.
{\it JWST} observations revealed broad components in H$\beta$ and H$\alpha$ from two objects, providing clear signatures of AGNs.
Two other objects observed with Keck/MOSFIRE show strong high-ionization lines, \ion{C}{4} $\lambda\lambda$1548, 1550 in one object \citep{onoue21} 
and \ion{N}{4} $\lambda\lambda$1483, 1487 in the other (M. Onoue et al., in prep.).\footnote{
The ionization potentials of \ion{C}{4} and \ion{N}{4} are 64 eV and 77 eV, respectively.}
Their large rest-frame EWs are difficult to explain with star-forming activity alone, pointing to the presence of hard AGN radiation.
The remaining three objects were observed with X-Shooter on the Very Large Telescope, but no emission lines other than Ly$\alpha$ were detected.
However, the sensitivity of the observations is somewhat lower than those mentioned above. 
Overall, accumulating evidence suggests that at least some of the SHELLQs objects with luminous and narrow Ly$\alpha$ host AGNs. 

\subsection{Extended emission and merging signatures}

\begin{figure*}
\epsscale{1.15}
\plotone{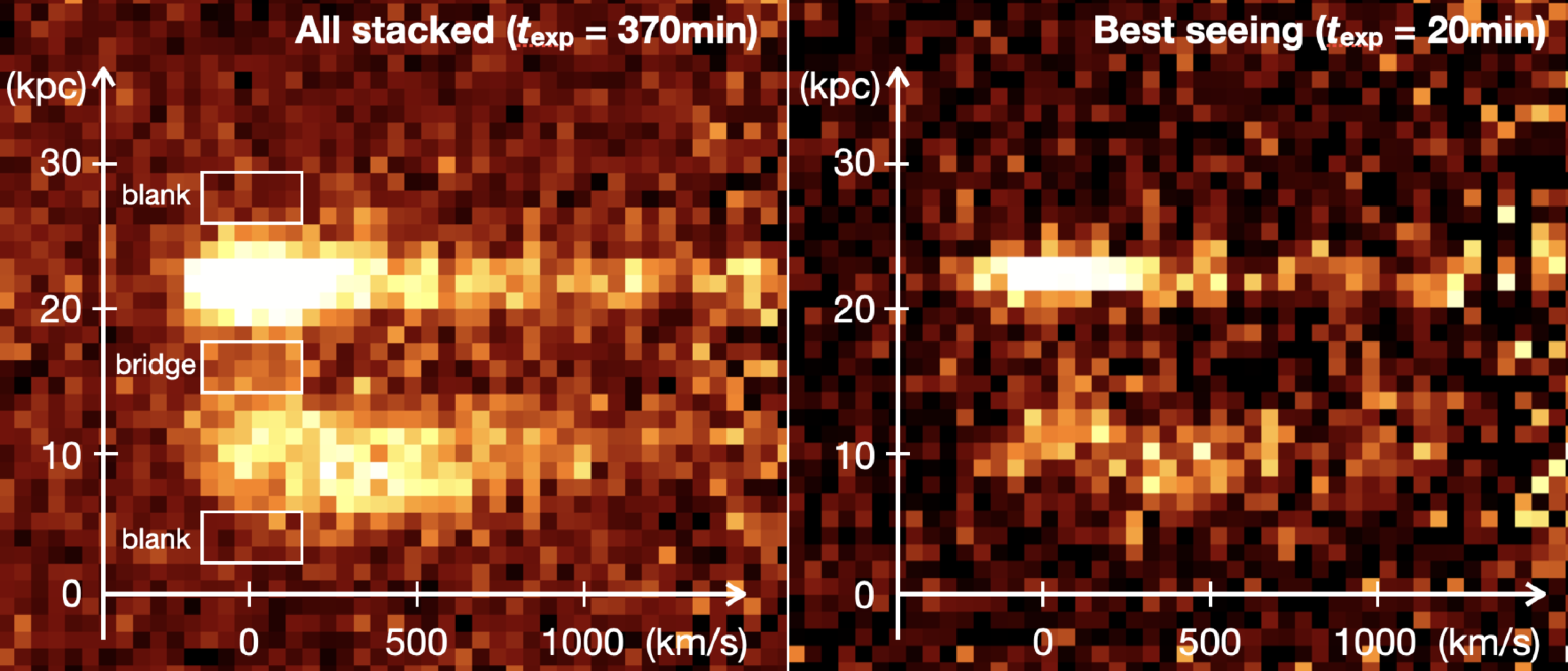}
\caption{Close-up view of the two-dimensional Ly$\alpha$ spectra of C1 (upper trace, at $\sim$22 kpc) and C2 (lower trace, at $\sim$10 kpc). 
The left panel displays the stacked data of all available exposures, totaling 370 min (i.e., an enlarged view of Figure \ref{fig:fc}).
The white boxes represent the regions used to calculate the S/N of the bridging emission (signal in the ``bridge" box and noise in the ``blank" boxes).
The right panel displays a single 20-min exposure taken under the best seeing condition ($\sim$0\arcsec4), which reveals a spatially resolved Ly$\alpha$ profile in C2.
The angular distance of 1\arcsec\ corresponds to 5.7 kpc at the source redshift.
\label{fig:lya_structure}}
\end{figure*}

The two objects are likely in physical association with each other, as indicated by their close separation in both the transverse and line-of-sight directions.
The angular separation of 2\arcsec.0 corresponds to a projected distance of 12 kpc (proper) or 82 kpc (comoving).
Moreover, we see extended Ly$\alpha$ emission bridging the two objects, as is clear from Figure \ref{fig:lya_structure} (left).
This emission component is detected with the signal-to-noise ratio (S/N) of $\sim$8, when the signal and noise are measured in the ``bridge" and ``blank" boxes indicated in the figure, respectively.
More spectacular bridging, tails, and other extended structures have been identified around C1 and C2 in the [\ion{C}{2}] 158$\mu$m line emission with
our ALMA observations, whose analysis will be presented in a companion paper (T. Izumi et al., in prep.).

\begin{figure}
\epsscale{1.15}
\plotone{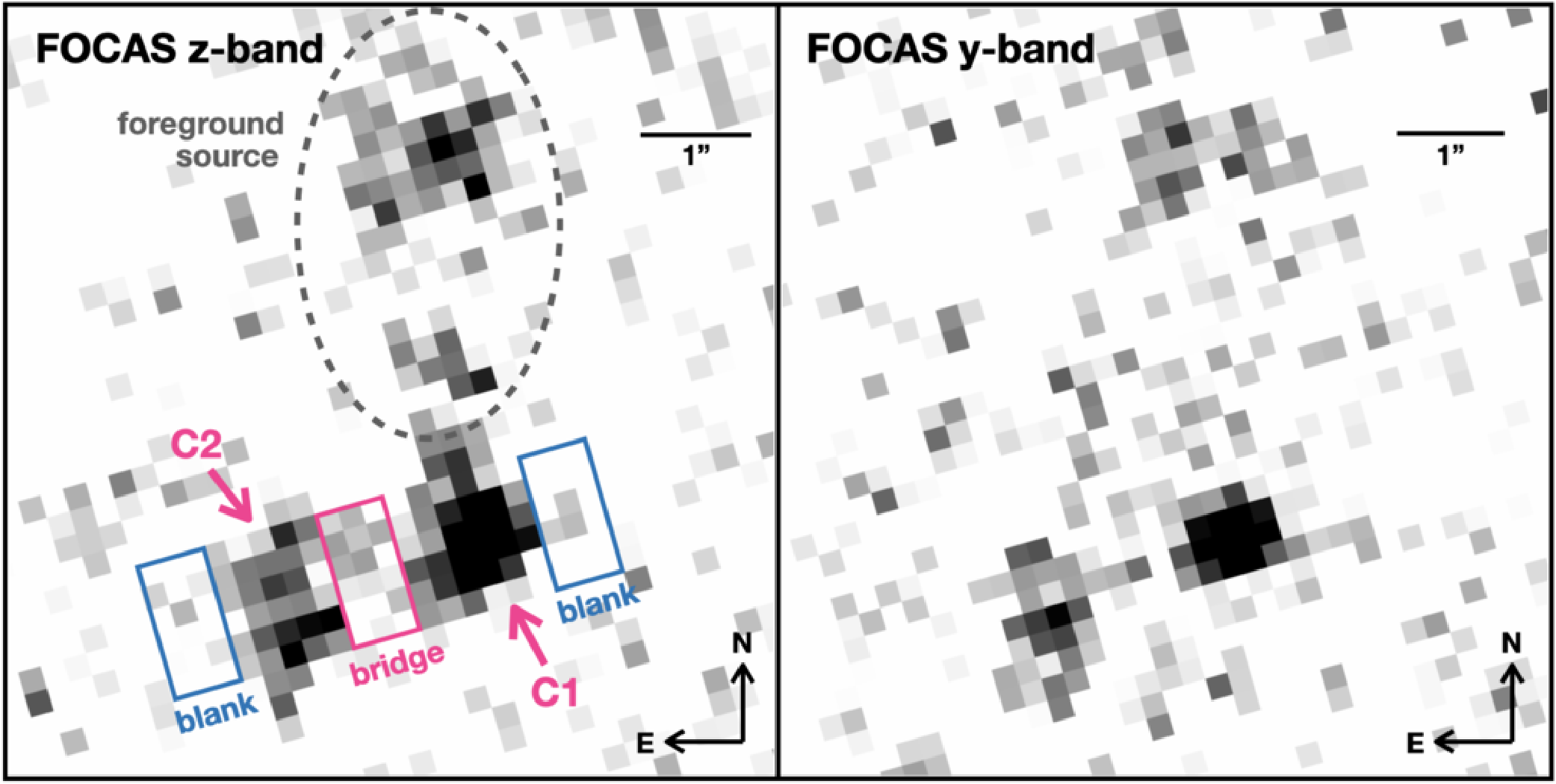}
\caption{FOCAS $z$-band (left) and $y$-band (right) images around the two quasars, obtained under seeing conditions of 0\arcsec.4--0\arcsec.6. 
North is up and East to the left.
The scale bars represent 1\arcsec.
The boxes represent the regions used to calculate the S/N of the bridging emission (signal in the ``bridge" box and noise in the ``blank" boxes).
\label{fig:focasIM}}
\end{figure}

The rest-UV emission connecting the two sources is also visible in our deep FOCAS images presented in Figure \ref{fig:focasIM}.
The bridging emission was detected only in the $z$ band containing Ly$\alpha$, with S/N $\sim$ 5 when the signal and noise are measured in the ``bridge" and ``blank" boxes indicated in the figure.
The $z$- and $y$-band images were taken on May 10 -- 11, 2019, under the seeing condition of 0\arcsec.4--0\arcsec.6.
The total exposure time is 30 min in each filter.



Similar extended rest-UV emission is seen around other high-$z$ quasars \citep{farina19}, in some cases accompanied by a merging galaxy \citep{decarli19a}.
While such emission is sometimes observed around an isolated quasar, the fact that it is visible only in the area connecting C1 and C2 in the present case provides 
a strong indication of a merger in progress. 
If the nearly 1:1 ratio of black hole masses indicates similar stellar masses in the merging two quasars, then it agrees with the results from hydrodynamical simulations, which predict that dual SMBH activity appears most frequently in merging galaxies with a mass ratio close to one, in close separations \citep[$<$10 kpc; e.g.,][]{capelo15, capelo17}.

The above FOCAS images also confirmed the spatial extendedness of C2, which has the normalized second-order moments of 
$\mu_z = 1.92 \pm 0.25$ and $\mu_y = 1.63 \pm 0.21$.
The large $\mu_z$ is most likely due to the spatially resolved Ly$\alpha$ of C2 (see Figure \ref{fig:lya_structure}), while the measurement of $\mu_y > 1$ may suggest a contribution from the host galaxy to the continuum emission.
C1 has a more compact shape, with $\mu_z = 1.29 \pm 0.16$ and $\mu_y = 1.34 \pm 0.15$. 
On the other hand, 
the FOCAS spectra in Figure \ref{fig:spec1} show 
no evidence of interstellar absorption lines such as \ion{Si}{2} $\lambda$1260, \ion{Si}{2} $\lambda$1304, and \ion{C}{2} $\lambda$1335, indicating that the host galaxy makes a subdominant contribution at most to the continuum spectrum, in either C1 or C2.
We note that our SMBH mass estimates are not affected by host galaxy contamination, since they are derived from the luminosity and width of the broad Ly$\alpha$ component only.
The bolometric luminosity and Eddington ratios reported above would be upper limits, if there were significant host galaxy contamination.

We have ruled out the possibility that these two sources are gravitationally-lensed images of a single quasar.
The ALMA observations mentioned above revealed significantly brighter ($>$5 times) far-IR continuum emission from C1 than from C2, in contrast to their similar brightness in the rest-UV ($\sim$0.6 mag difference; see Table \ref{tab:measurements}).
A continuous velocity gradient is observed throughout C1, C2, and the surrounding extended [\ion{C}{2}] 158 $\mu$m emission (blueshift to redshift from East to West, overall). 
In addition, only C2 has a spatially resolved Ly$\alpha$ profile, in which the West and East edges have a relative velocity offset of $\sim$300 km s$^{-1}$
and the nucleus has a velocity component extending to $\sim$1000 km s$^{-1}$.
We will present detailed analysis of gas kinematics, combining both the Ly$\alpha$ and [\ion{C}{2}] measurements, in the companion paper.

\section{Summary \label{sec:summary}}

This letter is the twentieth in a series of publications from the SHELLQs project, a high-$z$ quasar survey based on HSC-SSP imaging.
We report the serendipitous discovery of two merging quasars at $z$ = 6.05, one of the most distant pairs of quasars or AGNs known to date.
The quasars, HSC $J$121503.42$-$014858.7 (C1) and HSC $J$121503.55$-$014859.3 (C2), have similar rest-UV properties overall, with
$M_{1450} = -23.106 \pm 0.017$ (C1) and $-22.662 \pm 0.024$ (C2).
Both sources have a broad Ly$\alpha$ component with FWHM $>$ 1000 km s$^{-1}$, giving crude estimates of SMBH masses of $\log (M_{\rm BH}/M_\odot) = 8.1 \pm 0.3$.
The close separation (2\arcsec.0, corresponding to a projected proper distance of 12 kpc) and the bridging emission structure indicate that the two objects are undergoing a merger, which may have caused the observed quasar activity.
Indeed, ALMA observations have revealed spectacular extended structure surrounding the two quasars, whose detailed analysis will be presented in our companion paper (T. Izumi et al., in prep.).


\acknowledgments

This research is based on data collected at Subaru Telescope, which is operated by the National Astronomical Observatory of Japan. 
We are honored and grateful for the opportunity of observing the Universe from Maunakea, which has the cultural, historical and natural significance in Hawaii.
We appreciate the staff members of the telescope for their support during our FOCAS observations.

This research is based in part on data obtained at the international Gemini Observatory, a program of NSF's NOIRLab, via the time exchange program between Gemini and the Subaru Telescope. 
The international Gemini Observatory at NOIRLab is managed by the Association of Universities for Research in Astronomy (AURA) under a cooperative agreement with the National Science Foundation on behalf of the Gemini partnership: 
the National Science Foundation (United States), 
the National Research Council (Canada), 
Agencia Nacional de Investigaci\'{o}n y Desarrollo (Chile), 
Ministerio de Ciencia, Tecnolog\'{i}a e Innovaci\'{o}n (Argentina), 
Minist\'{e}rio da Ci\^{e}ncia, Tecnologia, Inova\c{c}\~{o}es e Comunica\c{c}\~{o}es (Brazil), 
and Korea Astronomy and Space Science Institute (Republic of Korea).

Y. M. was supported by the Japan Society for the Promotion of Science (JSPS) KAKENHI Grant No. 21H04494. 
K. I. acknowledges the support under the grant PID2022-136827NB-C44 provided by MCIN/AEI /10.13039/501100011033 / FEDER, EU.

The HSC collaboration includes the astronomical communities of Japan and Taiwan, and Princeton University.  The HSC instrumentation and software were developed by the National Astronomical Observatory of Japan (NAOJ), the Kavli Institute for the Physics and Mathematics of the Universe (Kavli IPMU), the University of Tokyo, the High Energy Accelerator Research Organization (KEK), the Academia Sinica Institute for Astronomy and Astrophysics in Taiwan (ASIAA), and Princeton University.  Funding was contributed by the FIRST program from the Japanese Cabinet Office, the Ministry of Education, Culture, Sports, Science and Technology (MEXT), the Japan Society for the Promotion of Science (JSPS), Japan Science and Technology Agency  (JST), the Toray Science  Foundation, NAOJ, Kavli IPMU, KEK, ASIAA, and Princeton University.
 
This letter is based on data collected at the Subaru Telescope and retrieved from the HSC data archive system, which is operated by Subaru Telescope and Astronomy Data Center (ADC) at NAOJ. Data analysis was in part carried out with the cooperation of Center for Computational Astrophysics (CfCA) at NAOJ.  
 
This letter makes use of software developed for Vera C. Rubin Observatory. We thank the Rubin Observatory for making their code available as free software at http://pipelines.lsst.io/. 
 
The Pan-STARRS1 Surveys (PS1) and the PS1 public science archive have been made possible through contributions by the Institute for Astronomy, the University of Hawaii, the Pan-STARRS Project Office, the Max Planck Society and its participating institutes, the Max Planck Institute for Astronomy, Heidelberg, and the Max Planck Institute for Extraterrestrial Physics, Garching, The Johns Hopkins University, Durham University, the University of Edinburgh, the Queen's University Belfast, the Harvard-Smithsonian Center for Astrophysics, the Las Cumbres Observatory Global Telescope Network Incorporated, the National Central University of Taiwan, the Space Telescope Science Institute, the National Aeronautics and Space Administration under grant No. NNX08AR22G issued through the Planetary Science Division of the NASA Science Mission Directorate, the National Science Foundation grant No. AST-1238877, the University of Maryland, Eotvos Lorand University (ELTE), the Los Alamos National Laboratory, and the Gordon and Betty Moore Foundation.


\begin{thebibliography}{}
\bibitem[Abbott et al.(2018)]{abbott18} Abbott, T.~M.~C., Abdalla, F.~B., Allam, S., et al.\ 2018, \apjs, 239, 18. doi:10.3847/1538-4365/aae9f0
\bibitem[Aihara et al.(2018)]{aihara18} Aihara, H., Arimoto, N., Armstrong, R., et al.\ 2018, \pasj, 70, S4 
\bibitem[Bosch et al.(2018)]{bosch18} Bosch, J., Armstrong, R., Bickerton, S., et al.\ 2018, \pasj, 70, S5. doi:10.1093/pasj/psx080
\bibitem[Capelo et al.(2017)]{capelo17} Capelo, P.~R., Dotti, M., Volonteri, M., et al.\ 2017, \mnras, 469, 4437. doi:10.1093/mnras/stx1067
\bibitem[Capelo et al.(2015)]{capelo15} Capelo, P.~R., Volonteri, M., Dotti, M., et al.\ 2015, \mnras, 447, 2123. doi:10.1093/mnras/stu2500
\bibitem[Decarli et al.(2019a)]{decarli19a} Decarli, R., Dotti, M., Ba{\~n}ados, E., et al.\ 2019, \apj, 880, 157. doi:10.3847/1538-4357/ab297f
\bibitem[Decarli et al.(2019b)]{decarli19b} Decarli, R., Mignoli, M., Gilli, R., et al.\ 2019, \aap, 631, L10. doi:10.1051/0004-6361/201936813
\bibitem[Decarli et al.(2017)]{decarli17} Decarli, R., Walter, F., Venemans, B.~P., et al.\ 2017, \nat, 545, 457. doi:10.1038/nature22358
\bibitem[De Rosa et al.(2019)]{derosa19} De Rosa, A., Vignali, C., Bogdanovi{\'c}, T., et al.\ 2019, \nar, 86, 101525. doi:10.1016/j.newar.2020.101525
\bibitem[Elias et al.(2006)]{elias06} Elias, J.~H., Joyce, R.~R., Liang, M., et al.\ 2006, \procspie, 6269, 62694C 
\bibitem[Fan et al.(2023)]{fan23} Fan, X., Ba{\~n}ados, E., \& Simcoe, R.~A.\ 2023, \araa, 61, 373. doi:10.1146/annurev-astro-052920-102455
\bibitem[Farina et al.(2019)]{farina19} Farina, E.~P., Arrigoni-Battaia, F., Costa, T., et al.\ 2019, \apj, 887, 196. doi:10.3847/1538-4357/ab5847
\bibitem[Gaia Collaboration et al.(2016)]{gaia16} Gaia Collaboration, Prusti, T., de Bruijne, J.~H.~J., et al.\ 2016, \aap, 595, A1. doi:10.1051/0004-6361/201629272
\bibitem[Greene et al.(2023)]{greene23} Greene, J.~E., Labbe, I., Goulding, A.~D., et al.\ 2023, arXiv:2309.05714. doi:10.48550/arXiv.2309.05714
\bibitem[Gunn \& Peterson(1965)]{gunn65} Gunn, J.~E., \& Peterson, B.~A.\ 1965, \apj, 142, 1633 
\bibitem[Hao et al.(2005)]{hao05} Hao, L., Strauss, M.~A., Tremonti, C.~A., et al.\ 2005, \aj, 129, 1783. doi:10.1086/428485
\bibitem[Harikane et al.(2022)]{harikane22} Harikane, Y., Ono, Y., Ouchi, M., et al.\ 2022, \apjs, 259, 20. doi:10.3847/1538-4365/ac3dfc
\bibitem[Harikane et al.(2023)]{harikane23} Harikane, Y., Zhang, Y., Nakajima, K., et al.\ 2023, \apj, 959, 39. doi:10.3847/1538-4357/ad029e
\bibitem[Hopkins et al.(2006)]{hopkins06} Hopkins, P.~F., Hernquist, L., Cox, T.~J., et al.\ 2006, \apjs, 163, 1. doi:10.1086/499298
\bibitem[Inada et al.(2012)]{inada12} Inada, N., Oguri, M., Shin, M.-S., et al.\ 2012, \aj, 143, 119. doi:10.1088/0004-6256/143/5/119
\bibitem[Kashikawa et al.(2002)]{kashikawa02} Kashikawa, N., Aoki, K., Asai, R., et al.\ 2002, \pasj, 54, 819 
\bibitem[Konno et al.(2016)]{konno16} Konno, A., Ouchi, M., Nakajima, K., et al.\ 2016, \apj, 823, 20 
\bibitem[Kormendy \& Ho(2013)]{kormendy13} Kormendy, J. \& Ho, L.~C.\ 2013, \araa, 51, 511. doi:10.1146/annurev-astro-082708-101811
\bibitem[Maiolino et al.(2023)]{maiolino23} Maiolino, R., Scholtz, J., Curtis-Lake, E., et al.\ 2023, arXiv:2308.01230. doi:10.48550/arXiv.2308.01230
\bibitem[Matthee et al.(2023)]{matthee23} Matthee, J., Naidu, R.~P., Brammer, G., et al.\ 2023, arXiv:2306.05448. doi:10.48550/arXiv.2306.05448
\bibitem[Matsuoka et al.(2016)]{p1} Matsuoka, Y., Onoue, M., Kashikawa, N., et al.\ 2016, \apj, 828, 26
\bibitem[Matsuoka et al.(2018a)]{p4} Matsuoka, Y., Iwasawa, K., Onoue, M., et al.\ 2018a, \apjs, 237, 5
\bibitem[Matsuoka et al.(2018b)]{p2} Matsuoka, Y., Onoue, M., Kashikawa, N., et al.\ 2018b, \pasj, 70, S35
\bibitem[Matsuoka et al.(2018c)]{p5} Matsuoka, Y., Strauss, M.~A., Kashikawa, N., et al.\ 2018c, \apj, 869, 150 
\bibitem[Matsuoka et al.(2023)]{p19} Matsuoka, Y., Onoue, M., Iwasawa, K., et al.\ 2023, \apjl, 949, L42. doi:10.3847/2041-8213/acd69f
\bibitem[Matsuoka et al.(2022)]{p16} Matsuoka, Y., Iwasawa, K., Onoue, M., et al.\ 2022, \apjs, 259, 18. doi:10.3847/1538-4365/ac3d31
\bibitem[Matsuoka et al.(2019a)]{p10} Matsuoka, Y., Iwasawa, K., Onoue, M., et al.\ 2019, \apj, 883, 183. doi:10.3847/1538-4357/ab3c60
\bibitem[Matsuoka et al.(2019b)]{p7} Matsuoka, Y., Onoue, M., Kashikawa, N., et al.\ 2019, \apjl, 872, L2. doi:10.3847/2041-8213/ab0216
\bibitem[Newman et al.(2012)]{newman12} Newman, S.~F., Genzel, R., F{\"o}rster-Schreiber, N.~M., et al.\ 2012, \apj, 761, 43. doi:10.1088/0004-637X/761/1/43
\bibitem[Oke \& Gunn(1983)]{oke83} Oke, J.~B., \& Gunn, J.~E.\ 1983, \apj, 266, 713 
\bibitem[Onoue et al.(2019)]{onoue19} Onoue, M., Kashikawa, N., Matsuoka, Y., et al.\ 2019, \apj, 880, 77. doi:10.3847/1538-4357/ab29e9
\bibitem[Onoue et al.(2018)]{onoue18} Onoue, M., Kashikawa, N., Uchiyama, H., et al.\ 2018, \pasj, 70, S31. doi:10.1093/pasj/psx092
\bibitem[Onoue et al.(2021)]{onoue21} Onoue, M., Matsuoka, Y., Kashikawa, N., et al.\ 2021, \apj, 919, 61. doi:10.3847/1538-4357/ac0f07
\bibitem[P{\^a}ris et al.(2012)]{paris12} P{\^a}ris, I., Petitjean, P., Aubourg, {\'E}., et al.\ 2012, \aap, 548, A66. doi:10.1051/0004-6361/201220142
\bibitem[Rakshit et al.(2020)]{rakshit20} Rakshit, S., Stalin, C.~S., \& Kotilainen, J.\ 2020, \apjs, 249, 17. doi:10.3847/1538-4365/ab99c5
\bibitem[Richards et al.(2006)]{richards06} Richards, G.~T., Haiman, Z., Pindor, B., et al.\ 2006, \aj, 131, 49. doi:10.1086/498063
\bibitem[Rorai et al.(2017)]{rorai17} Rorai, A., Hennawi, J.~F., O{\~n}orbe, J., et al.\ 2017, Science, 356, 418. doi:10.1126/science.aaf9346
\bibitem[Sandoval et al.(2023)]{sandoval23} Sandoval, B., Foord, A., Allen, S.~W., et al.\ 2023, arXiv:2312.02311. doi:10.48550/arXiv.2312.02311
\bibitem[Schlegel et al.(1998)]{schlegel98} Schlegel, D.~J., Finkbeiner, D.~P., \& Davis, M.\ 1998, \apj, 500, 525 
\bibitem[Schindler et al.(2023)]{schindler23} Schindler, J.-T., Ba{\~n}ados, E., Connor, T., et al.\ 2023, \apj, 943, 67. doi:10.3847/1538-4357/aca7ca
\bibitem[Schneider et al.(2010)]{schneider10} Schneider, D.~P., Richards, G.~T., Hall, P.~B., et al.\ 2010, \aj, 139, 2360. doi:10.1088/0004-6256/139/6/2360
\bibitem[Shen et al.(2021)]{shen21} Shen, Y., Chen, Y.-C., Hwang, H.-C., et al.\ 2021, Nature Astronomy, 5, 569. doi:10.1038/s41550-021-01323-1
\bibitem[Shen et al.(2011)]{shen11} Shen, Y., Richards, G.~T., Strauss, M.~A., et al.\ 2011, \apjs, 194, 45. doi:10.1088/0067-0049/194/2/45
\bibitem[Silverman et al.(2020)]{silverman20} Silverman, J.~D., Tang, S., Lee, K.-G., et al.\ 2020, \apj, 899, 154. doi:10.3847/1538-4357/aba4a3
\bibitem[Sobral et al.(2018)]{sobral18} Sobral, D., Matthee, J., Darvish, B., et al.\ 2018, \mnras, 477, 2817. doi:10.1093/mnras/sty782
\bibitem[Spinoso et al.(2020)]{spinoso20} Spinoso, D., Orsi, A., L{\'o}pez-Sanjuan, C., et al.\ 2020, \aap, 643, A149. doi:10.1051/0004-6361/202038756
\bibitem[Swinbank et al.(2019)]{swinbank19} Swinbank, A.~M., Harrison, C.~M., Tiley, A.~L., et al.\ 2019, \mnras, 487, 381. doi:10.1093/mnras/stz1275
\bibitem[Takahashi et al.(2024)]{takahashi24} Takahashi, A., Matsuoka, Y., Onoue, M., et al.\ 2024, \apj, 960, 112. doi:10.3847/1538-4357/ad045e
\bibitem[Tang et al.(2021)]{tang21} Tang, S., Silverman, J.~D., Ding, X., et al.\ 2021, \apj, 922, 83. doi:10.3847/1538-4357/ac1ff0
\bibitem[{\"U}bler et al.(2023)]{ubler23} {\"U}bler, H., Maiolino, R., P{\'e}rez-Gonz{\'a}lez, P.~G., et al.\ 2023, arXiv:2312.03589. doi:10.48550/arXiv.2312.03589
\bibitem[Vanden Berk et al.(2001)]{vandenberk01} Vanden Berk, D.~E., Richards, G.~T., Bauer, A., et al.\ 2001, \aj, 122, 549 
\bibitem[Vestergaard \& Peterson(2006)]{vestergaard06} Vestergaard, M. \& Peterson, B.~M.\ 2006, \apj, 641, 689. doi:10.1086/500572
\bibitem[York et al.(2000)]{york00} York, D.~G., Adelman, J., Anderson, J.~E., et al.\ 2000, \aj, 120, 1579. doi:10.1086/301513
\bibitem[Yue et al.(2023)]{yue23} Yue, M., Fan, X., Yang, J., et al.\ 2023, \aj, 165, 191. doi:10.3847/1538-3881/acc2be
\bibitem[Yue et al.(2021)]{yue21} Yue, M., Fan, X., Yang, J., et al.\ 2021, \apjl, 921, L27. doi:10.3847/2041-8213/ac31a9
\end{thebibliography}
\end{document}